\documentclass[eqsecnum,amsmath,preprintnumbers,superscriptaddress,nofootinbib,aps,12pt]{revtex4} 

\pdfoutput=1

\usepackage{amssymb,amsfonts,multirow}

\usepackage{epsfig}

\usepackage{placeins}
\usepackage{xspace}
\usepackage{cancel}

\usepackage{amssymb,url}
\usepackage{graphicx}
\usepackage{hyperref}

\usepackage{color}

\usepackage{array}
\usepackage{amsmath}
\usepackage{slashed}

\definecolor{rossoCP3}{cmyk}{0,.88,.77,.40}

\long\def\del #1 \enddel { }

\usepackage{graphicx}
\usepackage{amsmath}
\usepackage{amssymb}
\usepackage{subfigure}

\usepackage{epsfig}

\usepackage{graphicx}
\usepackage{subfigure}
\usepackage{hyperref}

\def\beq{\begin{equation}}
\def\eeq{\end{equation}}

\def\bea{\arraycolsep .1em \begin{eqnarray}}
\def\eea{\end{eqnarray}}
\def\Tr{{\rm Tr}}

\def\eps{\epsilon}

\def\al#1{\alpha_{#1}}
\def\eq#1{(\ref{#1})}

\def\s0#1#2{\mbox{\small{$ \frac{#1}{#2} $}}}
\def\0#1#2{\frac{#1}{#2}}

\def\grgl{\:\hbox to -0.2pt{\lower2.5pt\hbox{$\sim$}\hss}{\raise3pt\hbox{$>$}}\:}
\def\klgl{\:\hbox to -0.2pt{\lower2.5pt\hbox{$\sim$}\hss}{\raise3pt\hbox{$<$}}\:}

\newcommand{\W}{\overline{W}}

\def\lsim{\mathrel{\rlap{\lower4pt\hbox{\hskip1pt$\sim$}}
    \raise1pt\hbox{$<$}}}                
\def\gsim{\mathrel{\rlap{\lower4pt\hbox{\hskip1pt$\sim$}}
    \raise1pt\hbox{$>$}}}                

\newcommand{\ea}[1]{
\begin{align}
#1
\end{align}
}

\begin{document}

\title{Vacuum stability of asymptotically safe gauge-Yukawa theories}
\author{Daniel~F.~Litim}
\email{d.litim@sussex.ac.uk}
\affiliation{\mbox{Department of Physics and Astronomy, University of  Sussex, Brighton, BN1 9QH, U.K.}}
\author{Matin~Mojaza}
\email{mojaza@nordita.org}
\affiliation{NORDITA, KTH Royal Institute of Technology and Stockholm University,\\
Roslagstullsbacken 23, SE-10691 Stockholm, Sweden}
\author{Francesco~Sannino}
\email{sannino@cp3-origins.net}
\affiliation{{\color{rossoCP3}CP${}^3$-Origins} \& the Danish Institute for Advanced Study, Danish IAS,
Univ.~of Southern Denmark, Campusvej 55, DK-5230 Odense}

\begin{abstract}
We study the phase diagram and the stability of the ground state  for certain four-dimensional gauge-Yukawa theories whose high-energy behaviour is controlled by an interacting fixed point.  We also provide analytical and numerical results for running couplings, their crossover scales, the separatrix, and the Coleman-Weinberg effective potential. Classical and quantum stability of the vacuum is established. 
\\[.7ex]
{\footnotesize Preprint: CP3-Origins-2014-049, DIAS-2014-049, NORDITA-2014-145}

\end{abstract}

\maketitle

\tableofcontents

\section{Introduction}

The recent discovery of asymptotically safe quantum field theories in four space-time dimensions   offers new directions for both fundamental 
theory and model building \cite{Litim:2014uca}. 
The main  novelty of these theories is the presence of an exact interacting ultraviolet (UV)  fixed point. This
distinguishes them 
from asymptotically free theories  where the UV fixed point is non-interacting \cite{Gross:1973id,Politzer:1973fx}.
Also, no additional symmetry principles such as space-time supersymmetry \cite{Bagger:1990qh} are required to ensure well-defined and predictive theories in the UV 
\cite{Litim:2011cp}. Instead, the fixed point arises dynamically through  renormalisable interactions between non-Abelian gauge fields, fermions, and scalars, and in a regime where asymptotic freedom is absent.
Crucially, the potentially dangerous growth of the gauge coupling towards the UV is countered by Yukawa interactions, while the notorious growth of Yukawa and scalar couplings is tamed by the fluctuations of  gauge and fermion fields. This has led to  theories with ``complete asymptotic safety'', meaning interacting UV  fixed points in all couplings \cite{Litim:2014uca}.
This is quite distinct  from 
the more conventional setup of
``complete asymptotic freedom'' \cite{Gross:1973ju,Cheng:1973nv,Callaway:1988ya}, where the UV dynamics of Yukawa and scalar interactions is  brought under control by asymptotically free gauge fields; see \cite{Holdom:2014hla,Giudice:2014tma} for recent studies.

In order to qualify as fundamental theories, UV  fixed points 
must  reside in the ``physical'' regime of parameter space
where the vacuum is stable quantum-mechanically. 
The central question we wish to address in this paper is:  Does the quantum vacuum remain stable at an interacting UV  fixed point?
It is well-known that fluctuations can induce the spontaneous breaking of symmetry  {\it \`a la}  Coleman and Weinberg \cite{Coleman:1973jx}. 
We will answer the question to the positive by performing a detailed analysis of the theory's phase diagram. We also offer  results for key features of the theory including characteristic energy scales, the running couplings, the separatrix, and the resummed Coleman-Weinberg potential.

The rest of the paper is organised as follows. We recall the main features of asymptotically safe gauge-Yukawa theories,  which is followed by a detailed study of the phase diagram and the UV completion of the theory including analytical expressions for running couplings along UV-safe trajectories (Sec.~\ref{SecPD}). We then perform a stability analysis of the scalar field potential, both classically and quantum-mechanically, also offering analytical and numerical results for the resummed Coleman-Weinberg effective potential (Sec.~\ref{CWA}), followed by our conclusions   (Sec.~\ref{Conclusions}).

\section{Phase diagram and UV completion}\label{SecPD}

In this section, after reviewing the set-up and salient perturbative properties of asymptotically safe gauge-Yukawa theories introduced in \cite{Litim:2014uca}, 
we discuss the phase diagram including the RG evolution of the quartic self-couplings. We also derive explicit  expressions for the UV-safe trajectories away from the UV stable fixed point including along the UV-IR connecting separatrices, and further quantities required for the subsequent stability analysis.

\subsection{Preliminaries}

\label{Theory}

Following \cite{Litim:2014uca}, we consider a massless quantum field theory with $SU(N_C)$ gauge fields $A^a_\mu$ and field strength $F^a_{\mu\nu}$ $(a=1,\cdots, N^2_C-1)$, $N_F$ flavors of fermions $Q_i$ $(i=1,\cdots,N_F)$ in the fundamental representation, and a $N_F\times N_F$ complex matrix scalar field $H$ invariant under $U(N_F)_L\times U(N_F)_R$ rotations and uncharged under the gauge group. The fundamental Lagrangian is given by the sum of the Yang-Mills term, the fermion and scalar kinetic terms, the Yukawa interaction, and scalar self-interaction terms,
\bea
\label{F2}
L&=& 
- \s0{1}{2} \Tr \,F^{\mu \nu} F_{\mu \nu}
+\Tr\left(
\overline{Q}\,  i\slashed{D}\, Q \right)
 +\Tr\,(\partial_\mu H ^\dagger\, \partial^\mu H) 
\nonumber 
\\
&&
+y \,\Tr\left(\overline{Q}_L H Q_R + \overline{Q}_R H^\dagger Q_L\right)
-u\,\Tr\,(H ^\dagger H )^2  
-v\,(\Tr\,H ^\dagger H )^2  \,,
\eea
where the decomposition $Q=Q_L+Q_R$ with $Q_{L/R}=\frac 12(1\pm \gamma_5)Q$ is understood. The trace $\Tr$ indicates the trace over both color and flavor indices.
The model has four classically marginal coupling constants in four space-time dimensions given by the gauge coupling $g$, the Yukawa coupling $y$, the quartic scalar couplings $h$ and the `double-trace' scalar coupling $v$,  which we write as
\beq\label{couplings}
\al g=\frac{g^2\,N_C}{(4\pi)^2}\,,\quad
\al y=\frac{y^{2}\,N_C}{(4\pi)^2}\,,\quad
\al h=\frac{{u}\,N_F}{(4\pi)^2}\,,\quad
\al v=\frac{{v}\,N^2_F}{(4\pi)^2}\,.
\eeq
We have also added the appropriate powers of $N_C$ and $N_F$ in the normalization of the couplings. The shorthand notation $\beta_i\equiv\partial_t\alpha_i$ with $i=(g,y,h,v)$ is employed to indicate the $\beta$-functions for the couplings \eq{couplings}. In the Veneziano limit where both $N_C$ and $N_F$ are large but their ratio fixed, the parameter
\begin{equation}\label{eps}
\epsilon=\frac{N_F}{N_C}-\frac{11}{2}
\end{equation}
can take any real value. 
In the large-$N$ limit, the perturbative renormalisation group equations for the couplings \eq{couplings} have been obtained in \cite{Antipin:2013pya} in dimensional regularisation, also using  the results \cite{Machacek:1983tz,Machacek:1983fi,Machacek:1984zw}. In terms of \eq{eps} they are given by 
\begin{eqnarray}
\label{betag}
\beta_g&=&
\frac{4}{3}\eps\,\alpha_g^2 
 + 
\left\{ \left(25+\frac{26}{3}\eps\right) \alpha_g
-2
\left(\frac{11}{2}+\eps\right)^2 \alpha_y
\right\} \alpha_g^2 
\\
&&
+ \left\{ \left(\frac{701}{6}+  \0{53}{3} \eps - \0{112}{27} \eps^2\right) \alpha_g^2 - 
   \0{27}{8} (11 + 2 \eps)^2 \alpha_g \alpha_y + \0{1}{4} (11 + 2 \eps)^2 (20 + 3 \eps) \alpha_y^2\right\} \alpha_g^2
   \nonumber\\
\beta_y&=&
\alpha_y\, \Big\{ (13 + 2 \eps) \,\alpha_y-6\,\alpha_g \Big\} \\ \nonumber
&&+\alpha_y 
\left\{
\0{20 \eps-93}{6}\alpha_g^2 
+  (49 + 8 \eps) \alpha_g \alpha_y
-    \left(\0{385}{8} + \0{23}{2} \eps + \0{\eps^2}{2}\right) \alpha_y^2 
-  (44 + 8 \eps) \alpha_y \alpha_h 
+ 4 \alpha_h^2\right\} 
\label{betay}
\\
\label{betah}
  \beta_h&=&
 -(11+ 2\eps) \,\alpha_y^2+4\alpha_h(\alpha_y+2\alpha_h)\,,\\
\label{betav}
  \beta_v&=&
12 \alpha_h^2  +4\al v \left(\alpha_v+ 4 \alpha_h+\alpha_y\right)\,.
\end{eqnarray}
 for $\beta_g,\beta_y,\beta_h$ and $\beta_v$ up to $(3,2,1,1)$-loop order, respectively. In the terminology of  \cite{Litim:2014uca} we refer to this as the next-to-next-to-leading order (NNLO) approximation. The NLO approximation corresponds to the approximation in which the $(2,1,0,0)$-loop terms for $\beta_g,\beta_y,\beta_h$ and $\beta_v$ are retained. As discussed in \cite{Litim:2014uca}, this ordering of perturbation theory is also favoured by Weyl consistency conditions  \cite{Jack:1990eb,Antipin:2013pya,Jack:2014pua}. Additionally, we  consider the NLO${}^\prime$ approximation  in which the $(2,1,1,1)$-loop terms are retained (see Tab.~\ref{Tab}). We stress that the dynamics of the scalar couplings, central for our study of vacuum stability, arises for the first time within the NLO${}^\prime$ and NNLO approximations. 
 
The theory \eq{F2} is renormalisable within perturbation theory. For $\eps<0$ it also displays asymptotic freedom in the gauge sector. When $\eps$ is positive, asymptotic freedom is lost and the gauge sector becomes QED-like.  In \cite{Litim:2014uca}, it has been established that the theory \eq{F2} develops an exact, interacting UV fixed point in all four couplings, strictly controlled by perturbation theory, provided the parameter \eq{eps} is chosen to be
\begin{equation}\label{small}
0<\eps\ll 1\,.
\end{equation}
The existence of an interacting UV fixed point  ensures that the theory \eq{F2} remains predictive to highest energies at the fixed point. Another feature of this theory is that the scalar sector avoids a triviality bound due to residual interactions in the UV, meaning that the scalars can be viewed as elementary.

\begin{center}
\begin{table}
\begin{tabular}{c|cccc}
 \hline\hline
 {}\ \ \ coupling\ \ \ &\multicolumn{4}{c}{{}\ \ \ order in perturbation theory\ \ \ }\\ \hline
 $\alpha_g$&1&2&2&3 \\
  $\alpha_y$ &0&1&1&2 \\
   $\alpha_h$ &0&0&1&1\\
    $\alpha_v$ &0&0&1&1 \\ \hline
    \ \ approximation level\ \ &{}\ \ \ \ \ LO\ \ \ \ \  &\ \ NLO\ \  &\ \ NLO${}^\prime$\ \   &\ \ NNLO\ \ \\ \hline\hline
\end{tabular}
\caption{\label{Tab} Relation between approximation level and the loop order to which couplings are retained in perturbation theory.}
\end{table} 
\end{center}

\subsection{Fixed points}
In  \cite{Litim:2014uca}, it has been shown that the gauge-matter system \eq{betag} -- \eq{betav} may display three different ultraviolet fixed points in the regime \eq{small}. 
In an expansion in the small parameter $\eps$, the first fixed point has the coordinates  
\beq\label{alphaNNLO}
\begin{array}{rcl}
\alpha_g^*&=&
\frac{26}{57}\,\eps+ \frac{23 (75245 - 13068 \sqrt{23})}{370386}\,\eps^2
+{\cal O}(\eps^3)
\\[1.ex]
\alpha_y^*&=&
\frac{4}{19}\,\eps+\left(\frac{43549}{20577} - \frac{2300 \sqrt{23}}{6859}\right)\,\eps^2
+{\cal O}(\eps^3)
\\[1ex]
\alpha_h^*&=&
\frac{\sqrt{23}-1}{19}\,\eps+{\cal O}(\eps^2)
\,,\\[1.ex]
\al {v1}^*&=&
-\frac{1}{19} (2 \sqrt{23}-\sqrt{20 + 6 \sqrt{23}})\,\eps
+{\cal O}(\eps^2)\,.
\end{array}
\eeq
We refer to the fixed point \eq{alphaNNLO} as FP${}_1$. There are two more fixed points in the perturbative domain \eq{small}, a second fully interacting fixed point FP${}_2$ and a partial fixed point  FP${}_3$.  The coordinates of FP${}_2$ differ from FP${}_1$ in \eq{alphaNNLO}  in the coordinate $\alpha_{v}$, whose fixed point  becomes 
\beq
\label{FP2}
\begin{array}{rcl}
\alpha_{v2}^*&=&-\frac{1}{19} (2 \sqrt{23}+\sqrt{20 + 6 \sqrt{23}})\eps+{\cal O}(\eps^2)
\end{array}
\eeq 
instead. The second fixed point comes about because  the double trace coupling does not couple to the RG flow of $(\al g,\al y,\al h)$, and its own RG flow \eq{betav} is quadratic in $\alpha_v$ to all orders in $1/N_F$ and $1/N_C$. Therefore, $\al v$ may even interpolate between these two fixed points without affecting the fixed point  for $(\al g,\al y,\al h)$. 

Finally, a partial fixed point FP${}_3$ arises where all but the double-trace coupling  $\alpha_v$ settle onto a fixed point within the perturbative domain. The double-trace coupling continues to evolve logarithmically towards a perturbative Landau pole. Furthermore, $\alpha_h$  settles at a negative fixed point value while $\alpha_g^*$ and $\alpha_y^*$ remain equal to their values at \eq{alphaNNLO}  to this order in $\epsilon$. Towards UV momentum scales, the partial fixed point FP${}_3$ is then characterised by
\beq
\label{FP3}
\begin{array}{rcl}
\alpha^*_{h2}&=&
\displaystyle 
-(\sqrt{23}+1)\
\0\eps{19}+{\cal O}(\eps^2)\\[2ex]
\alpha_v(\mu)&=&
\displaystyle
\alpha_v(\mu_0)
+\frac {\eps}{4}\,\kappa\tan \left(\kappa\,\eps\ln\frac{\mu}{\mu_0}\right)
\end{array}
\eeq
where 
$\kappa=\frac{4}{19}\sqrt{6\sqrt{23}-20}\approx 0.624$. We have chosen the scale $\mu_0$ as a free parameter with $\alpha_v(\mu_0)=(2\,\eps\,\sqrt{23})/19\approx 0.505\,\eps$. The running for $\alpha_v$ in \eq{FP3} is an exact solution of its beta function \eq{betav} provided that $\alpha_g$, $\alpha_y$ and $\alpha_h$ have settled on FP${}_3$ to leading order in $\eps$.
From \eq{FP3} we observe  that $\alpha_v$  reaches its perturbative UV Landau pole at
\beq
\Lambda_{\rm Landau}\approx\mu_0\,\exp\frac{\pi}{2\,\kappa\,\eps}\gg \mu_0\,.
\eeq
A similar Landau pole is reached towards low energies.

\begin{figure}[t]
\centering
\includegraphics[width=.7\textwidth]{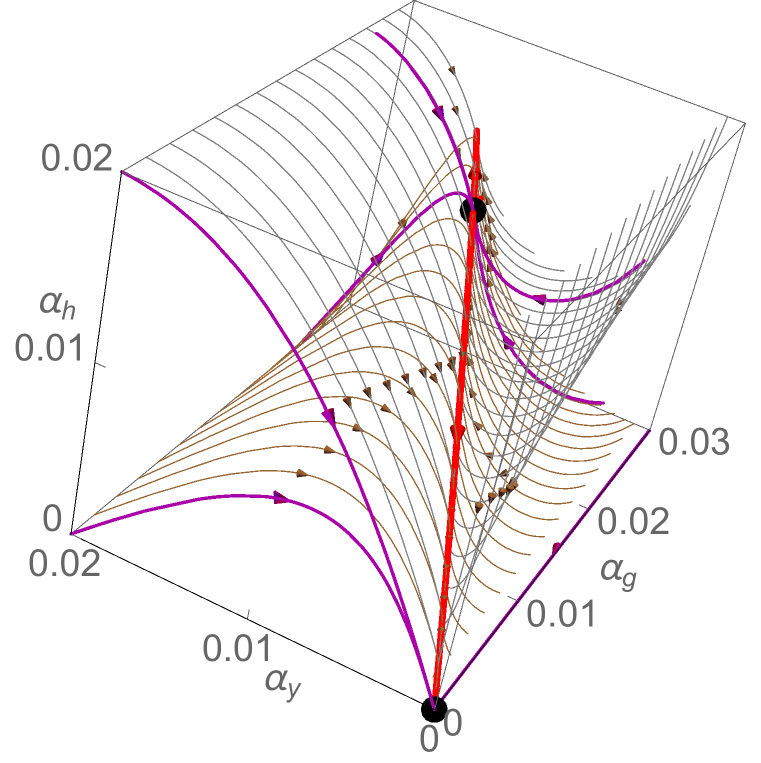}
\caption{\label{pPD}  The phase diagram 
in the gauge-Yukawa-scalar subsector  of couplings $(\al g,\al y,\al h)$ at NNLO  accuracy with $\eps=0.05$. Also shown are the UV and IR fixed points (dots), the UV safe trajectories (thick red line),  a few distinguished trajectories (thin magenta lines) and sets of generic trajectories (thin gray lines). Arrows point towards the IR (see main text).}
\end{figure}

\subsection{Phase diagram}
\begin{figure*}[t]
\begin{center}
\includegraphics[width=.471\textwidth]{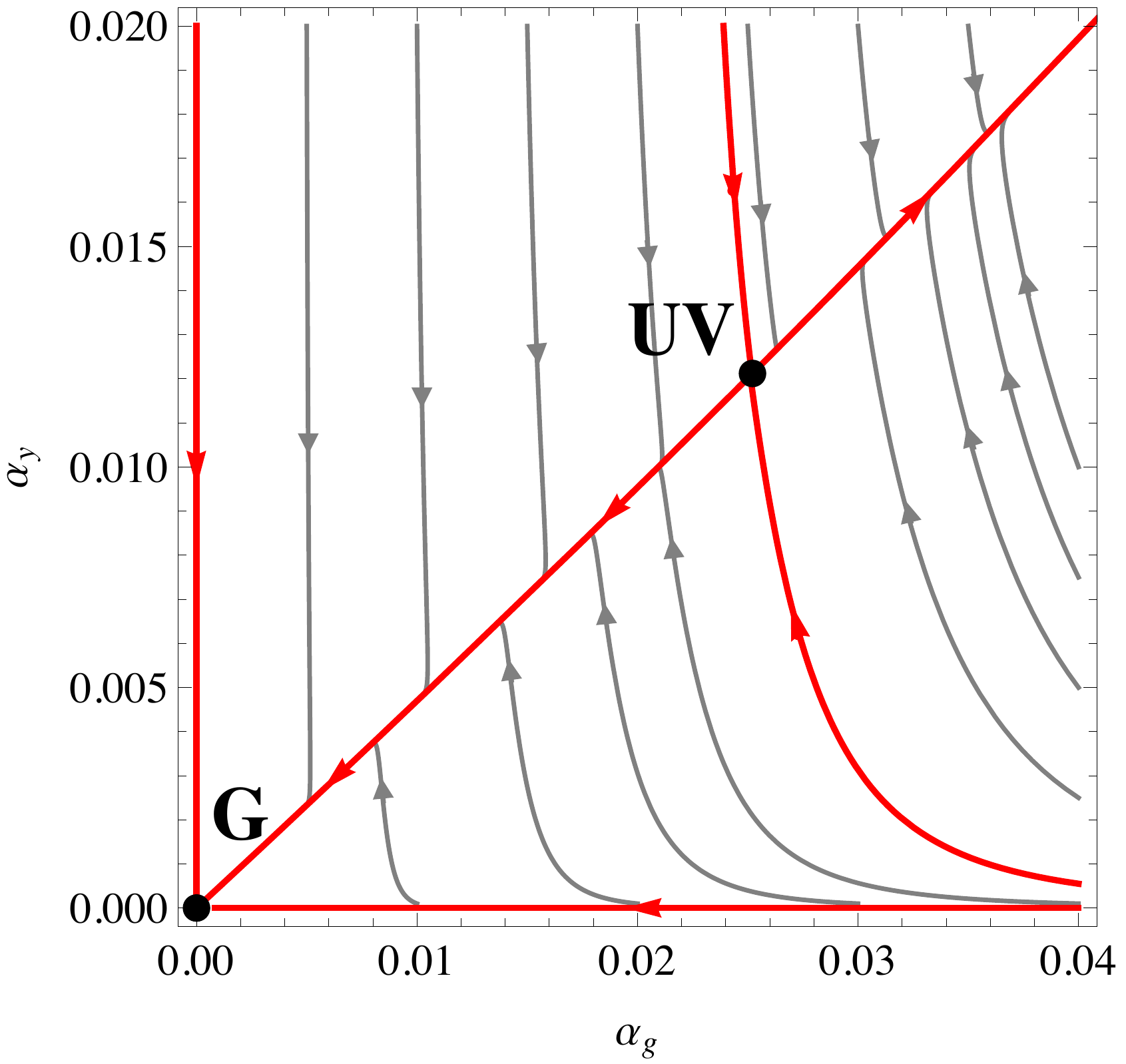}
\includegraphics[width=.47\textwidth]{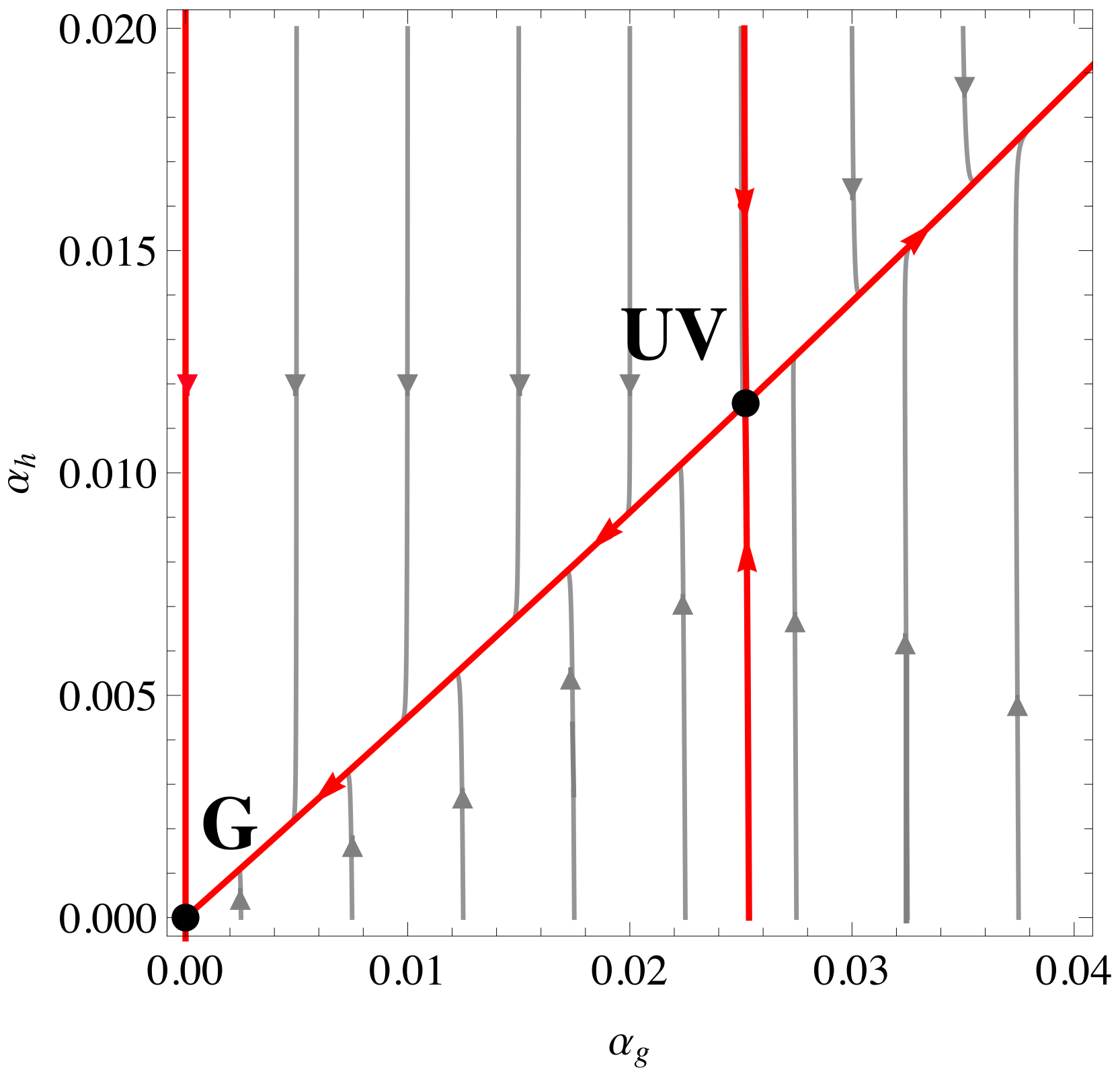}
\caption{\label{pPDproject}  The phase diagram of the gauge-Yukawa theory in the vicinity of the UV fixed point  at NNLO accuracy with $\eps=0.05$, projected onto the $(\al g,\al y)$ plane (left panel) and  the $(\al g,\al h)$ plane (right panel). 
 Shown are
the asymptotically safe fixed point (UV), the Gaussian fixed point (G), and various sample trajectories. RG trajectories point towards the IR.}
\label{pPhase}
\end{center}
\end{figure*}

\begin{figure*}[t]
\begin{center}
\includegraphics[width=.47\textwidth]{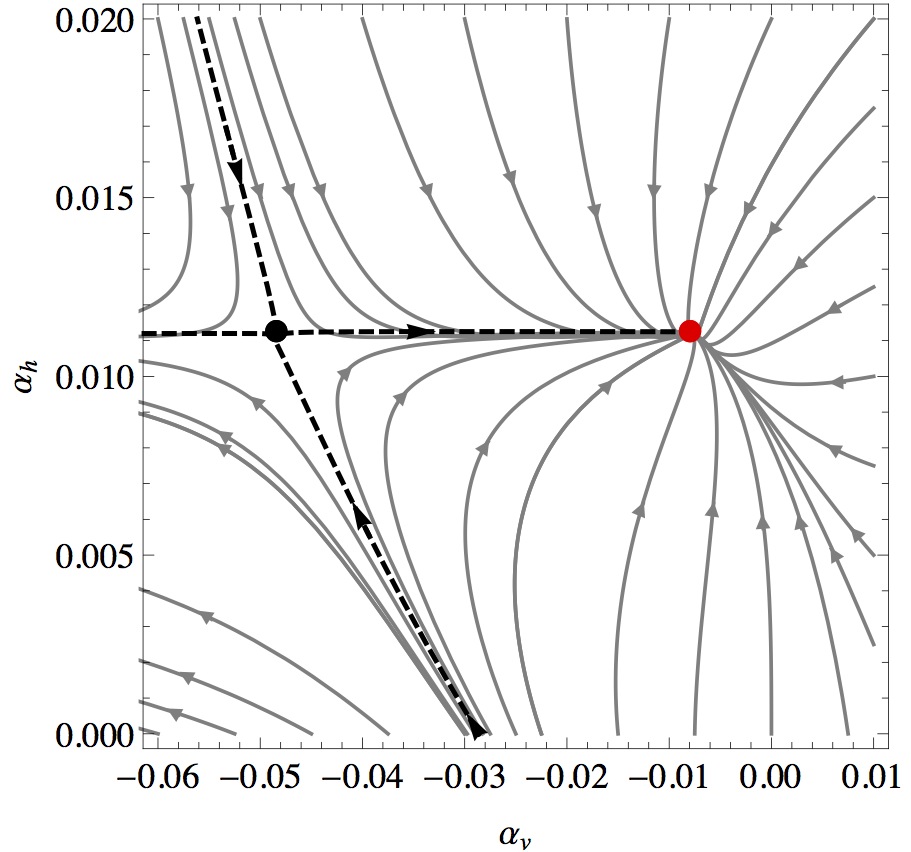}
\includegraphics[width=.47\textwidth]{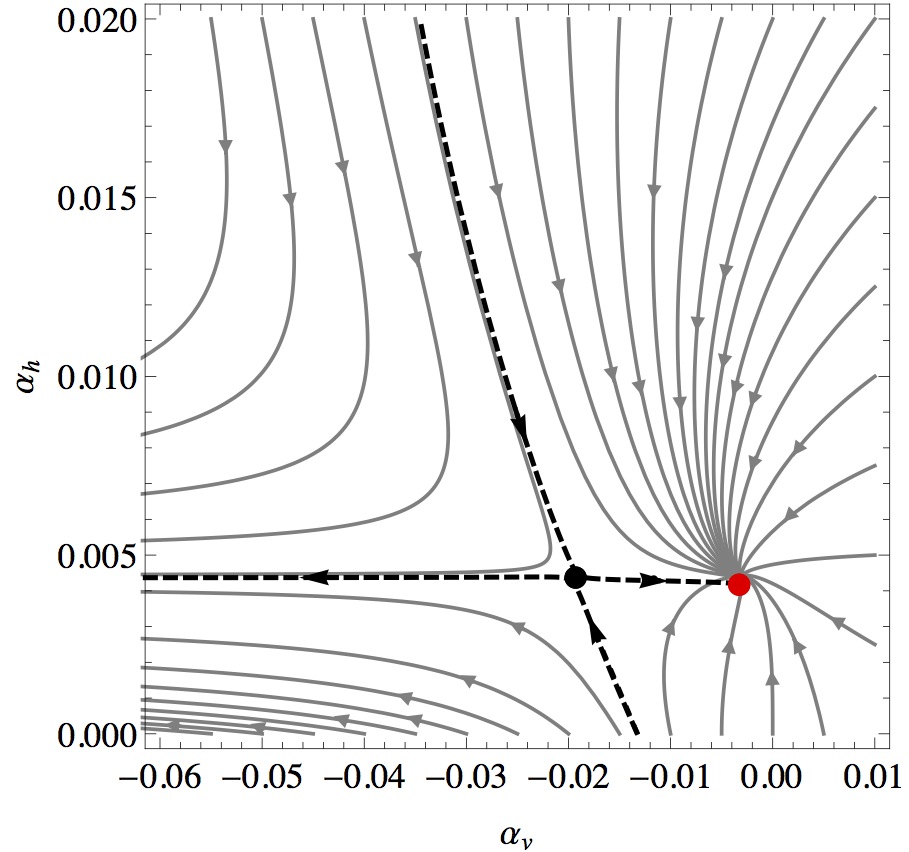}
\caption{\label{pPDscalar}  Projection of the phase diagram of the gauge-Yukawa theory onto the subspace of scalar couplings  (NNLO with $\eps=0.05$) with $(\al g,\al y)$ taking values on the UV-IR connecting separatrix (left panel: $\al g\approx 0.999\, \al g^*$  right panel: $\al g \approx0.397\, \al g^*$).
Red  (black) dots indicate the trajectory which connect the physical fixed point FP${}_1$ (unphysical fixed point FP${}_2$) with the Gaussian fixed point. In the scalar subsystem where the RG flow is parametrically faster by $1/\eps$, the separatrices appear as pseudo fixed points (see main text).}
\label{pPhase}
\end{center}
\end{figure*}

In \cite{Litim:2014uca}, the phase diagram of the model has been discussed at NLO accuracy. Here we extend the study to NLO${}^\prime$ and NNLO order including effects due to the running scalar couplings, and illustrate results in  Figs.~\ref{pPD},~\ref{pPDproject} and~\ref{pPDscalar} which show portions of the phase diagram in the vicinity of the UV and IR fixed points. 

Specifically, 
Fig.~\ref{pPD} shows the RG trajectories for the couplings $(\al g,\al y,\al h)$, including the UV and IR fixed points and the UV-complete trajectory (thick red line) connecting them. It corresponds to a weakly coupled theory in the IR limit. The second UV complete trajectory runs towards large couplings, leading to a strongly coupled low energy theory with chiral symmetry breaking or conformality in the IR. The thin (magenta) trajectories emanating out of the UV stable fixed point correspond to the two irrelevant eigendirections of the fixed point.  The other (thin magenta) trajectories illustrate that the Gaussian fixed point is an infrared fixed point for all couplings. Unlike in asymptotically free theories, it does not qualify as a UV fixed point. Finally, the thin (gray) lines show same trajectories which do not originate from a perturbative UV fixed point, but which nevertheless lead to a well-behaved weakly coupled theory at low energies. \\

Figs.~\ref{pPDproject} show the projection of Fig.~\ref{pPD} onto the $(\al g,\al y)$ and the $(\al g,\al h)$ plane, respectively. Notice  the near-perpendicular behaviour of trajectories close to the separatrix in the $(\al g,\al y)$ plane (left panel). This is  due to the fact that the velocity $\mu\partial_\mu$ of the RG flow of couplings along the separatrix is slower  by an order in $\eps\ll 1$. In the range of parameters considered here, the RG flow has a velocity of order $\epsilon$ set by the Yukawa and scalar quartic couplings, yet the flow of the gauge coupling is smaller by a power in $\eps$. However, as soon as couplings are very close to the UV-IR connecting trajectory, the flow velocity becomes of order $\eps^2$ for all couplings, controlled by the gauge coupling.  This effect is even more pronounced in the $(\al g,\al h)$ plane (right panel), because the scalar and gauge $\beta$-function are independent of each other at NNLO order. Consequently, the larger RG velocity of the scalar coupling becomes visible in the entire phase diagram, except along the separatrix.
This effect plays also a r\^ole in the scalar subsector $(\al h,\al v)$ for which  two snapshots are shown in Fig.~\ref{pPDscalar}. Here, the gauge and Yukawa couplings have been given values on the UV-IR connecting separatrix. Specifically, we have taken $\al g\approx 0.999\, \al g^*$ (left panel) and $\al g \approx0.397\, \al g^*$ (right panel), respectively, and $\al y$ determined via  \eq{critsurf}. Once again, in these projections, the  RG flows in the scalar subsector have  velocities of order $\eps$, whereas the flow velocity along the separatrix is of order $\eps^2$ and hence parametrically slower. Therefore the separatrix appears as a ``pseudo'' fixed point in this projection, indicated by a red dot. All other trajectories are attracted towards the separatrix. We also observe a second ``pseudo'' fixed point, indicated by a black dot. It relates to the trajectory which connects FP${}_2$ with the Gaussian fixed point. However, along this trajectory the scalar field potential is unbounded from below and the theory is not considered to be physical.

\subsection{UV completion}

Trajectories which emanate out of the UV stable fixed point correspond to finite theories at high energies. In the present case, the set of UV stable trajectories, sometimes denoted as the ``UV critical surface'', is found to be one-dimensional \cite{Litim:2014uca}. The reason for this is a strictly dynamical one: the fluctuations at the UV fixed point are such that only one linear combination of the four classically marginal couplings \eq{couplings} becomes quantum-mechanically relevant. 
Similar dynamical constraints can also arise for theories with ``complete asymptotic freedom'' \cite{Callaway:1988ya}. 
 More specifically, here, the one-dimensional critical surface leads to two UV safe trajectories. As indicated earlier, in the low-energy limit, they correspond to a weakly interacting theory of gluons, fermions and scalars with Gaussian scaling, and a strongly interacting theory with confinement, chiral symmetry breaking, or conformality.

The separatrix  connects the UV fixed point with the Gaussian one. It coincides with the UV critical surface close to the fixed point \cite{Litim:2014uca}, and  is characterised by global aspects of the phase diagram, namely the location of its fixed points and the eigendirections. Therefore, in general, identifying a separatrix requires global information about the phase diagram. For generic values of the parameter $\eps$, the separatrix can always be found numerically.  
For some considerations, it is useful to have explicit analytical expressions for the UV-IR connecting trajectory.
  Here, we explain how an analytical expression for the separatrix can be obtained locally to leading order in $\eps\ll 1$ using the following observations. 
For small $\eps\ll 1$, the four eigenvalues at the UV fixed point are given by
\beq\label{thetaNNLO}
\begin{array}{rcl}
\vartheta_1&=&
-\frac{104}{171}\,\eps^2
+{\cal O}(\eps^3)\\[2ex]
\vartheta_2&=&
\ \ \,\frac{52}{19}\, \eps 
+{\cal O}(\eps^2)\\[2ex]
\vartheta_3&=&
\ \ \frac{16\sqrt{23}}{19}\,\eps
+{\cal O}(\eps^2)
\\[2ex]
\vartheta_4&=&\ \  \frac{8}{19} \sqrt{20 + 6 \sqrt{23}}\,\eps+{\cal O}(\eps^2)\,.
\end{array}
\eeq
We note that the irrelevant (positive) eigenvalues are of order $\eps$ whereas the relevant (negative) eigenvalue is of order  $\eps^2$. In consequence, the velocity $\mu\partial_\mu$  of the RG evolution of couplings along the separatrix is $\propto \eps^2$, whereas the velocity of the RG evolution towards the separatrix is  $\propto \eps$. In the regime \eq{small} we conclude that the approach towards the separatrix is substantially faster than the evolution along the separatrix.
For fixed $\alpha_g$, this means that the corresponding value for the Yukawa coupling $\alpha_y$ on the separatrix can be found by solving $\beta_y=0$, to leading order in $\eps$. Following the same reasoning, for given $(\alpha_g,\alpha_y$) on the separatrix, the corresponding value for $\alpha_h$ is then found from solving $\beta_h=0$. Finally, $\alpha_v$ follows from solving $\beta_v=0$ after inserting $(\alpha_y,\alpha_h$) on the separatrix. Following this logic, we are led to solve
\beq\label{separatrixLO}
\begin{array}{rcl}
\beta_y&=&0\,,\\
\beta_h&=&0\,,\\
\beta_v&=&0
\end{array}
\eeq
for $(\al y,\al h,\al v)$ in terms of $\al g$ and to leading order in $\eps$. This leads to the following relations amongst couplings along the UV stable trajectories
\beq
\label{critsurf}
\begin{array}{rcl}
\al y &=&
 \frac{6}{13}\, \al g \ , \\[2ex]
\al h & = &
\frac{3}{26} (\sqrt{23}-1)\,\al g \ , \\[2ex]
\al v & = &
\frac{3}{26} (\sqrt{20 + 6\sqrt{23}}-2\sqrt{23})\,\al g \,.
\end{array}
\eeq
These relations for the separatrices are accurate in the limit $\eps\to 0$ \eq{small}. It remains to specify the RG running of the gauge coupling.

\subsection{Effective gauge coupling}\label{eff}
We now exploit the condition \eq{separatrixLO} to find 
the effective RG running of the gauge coupling along the separatrices. To simplify some of the subsequent expressions, we introduce $\alpha\equiv \alpha_g$ throughout this section. Its RG flow is given by
\begin{equation}\label{dalpha2}
 \partial_t\, \alpha=-B\,\alpha^2+C\,\alpha^3+{\cal O}(\alpha^4)\,.
\end{equation}
Below we will derive exact results valid for theories with \eq{dalpha2} and generic $B,C\neq 0$. For the theory at hand, we concentrate on parameters in the regime $B<0$ and $C<0$ which take the specific values
\beq\label{BC}
\begin{array}{rcl}
B&=&
\displaystyle
-\043 \eps\\[1.5ex]
C&=&
\displaystyle
-\frac{2}{3}\,\0{57-46\eps-8\eps^2}{13+\eps}\,.
\end{array}
\eeq
They arise from \eq{betag} by expressing all $\alpha_i$ $(i=y,h,v)$ in terms of $\alpha$, which are found by solving each of \eq{separatrixLO} to one-loop accuracy. Evidently, the effective gauge $\beta$-function along the separatrix \eq{dalpha2} displays three fixed points, a doubly-degenerated one at $\alpha_*=0$, and an interacting one at
\begin{equation}\label{fixed point2}
\alpha_*=B/C>0\,.
\end{equation}
Notice that in the absence of scalar fields and Yukawa interactions the coefficient $C$ would read $C=25>0$, and consequently the gauge coupling would not aquire an interacting UV fixed point in the physical domain $\alpha>0$. Including Yukawa interactions, the non-trivial fixed point is perturbative as long as $0\le\alpha_*\ll 1$. In terms of \eq{fixed point2}, the  solution to \eq{dalpha2} is 
\beq\label{solution2loop}
\left(
\frac{\mu}{\mu_0}
\right)^{-B\cdot\alpha_*}
=
\left(\frac{\al *-\alpha_0}{\al *-\alpha}\cdot\frac{\alpha}{\alpha_0}\right)
\exp \left(\frac{\al *}{\alpha_0}-\frac{\al *}{\alpha}\right)
\eeq
Here, $-B\cdot\alpha_*>0$,  and $\alpha\equiv\alpha(\mu)$ denotes the running gauge coupling along the separatrix, $\mu_0$  the initial scale, and $\alpha_0\equiv \alpha(\mu=\mu_0)$ the gauge coupling at that scale.
The solution \eq{solution2loop} provides a closed expression for $\mu=\mu(\alpha)$. One easily confirms that \eq{solution2loop} interpolates between the UV and IR fixed points. For $\alpha\to 0$ the exponential term dominates, thereby showing  that it corresponds to an IR fixed point with $\mu\to 0$. In return, $\alpha\to\alpha_*$ corresponds to the UV limit where the algebraic term in \eq{solution2loop} dominates, $\mu\to\infty$, and we recognize the appearance of the scaling exponent $\vartheta_1=B\cdot\al *$. Note that if $\alpha_0>\alpha_*$ the Gaussian fixed point cannot be reached. Rather, RG trajectories then evolve towards strong coupling for all scales $\mu<\mu_0$. 
In order to invert the solution  $\mu=\mu(\alpha)$ into $\alpha=\alpha(\mu)$ we introduce
\bea
\label{W}
W&=&\frac{\alpha_*}{\alpha}-1\\
\label{z}
z&=&
\left(\frac{\mu_0}{\mu}\right)^{-B\cdot\alpha_*}
\left(\frac{\alpha_*}{\alpha_0}-1\right)
\exp\left(\frac{\alpha_*}{\alpha_0}-1\right)\,.
\eea
For $\mu/\mu_0$ ranging between $[0,\infty]$ and $0<\alpha_0<\alpha_*$, we have that both $z$ and $W$ interpolate monotonously between $[\infty,0]$. For $0<\alpha_*<\alpha_0$, we have that $z$ becomes negative. Using these expressions, \eq{solution2loop} turns into
\beq z=W\exp W
\eeq
whose solution $W=W[z]$ is given by Lambert's $W$ function. Uniqueness is observed provided that $z\ge 0$, which always holds true for the separatrix towards the Gaussian. In this regime the Lambert W function is monotonically increasing, polynomially so for small $z$ and approaching $\ln z$ from below for large $z$. For negative $z$, the Lambert W function is multivalued. The relevant branch is then the one with $W[z]\to0$ for $z\to 0$. 

\begin{figure*}[t]
\begin{center}
\includegraphics[width=.7\textwidth]{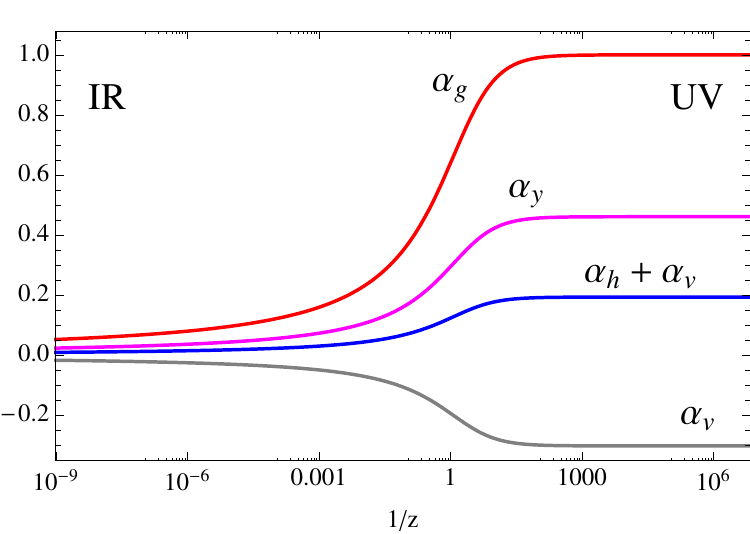}
\caption{\label{pRunAll} 
Renormalisation group running of all couplings along the UV-IR connecting separatrix \eq{critg}, \eq{critsurf} in the NLO${}^\prime$ approximation, with $z(\mu)$ defined in \eq{z}. All couplings are normalised in units of $\alpha_g^*$ ($\eps=0.05$). The UV-IR crossover takes place at $z(\mu)\approx 1$. Although $\al v$ stays negative throughout, we notice that the sum of quartic scalar coupings $\al h + \al v$ stays positive, approching zero from above in the deep IR.}
\label{pCW}
\end{center}
\end{figure*}

In terms of \eq{W} and \eq{z}, the fully resolved solution is finally written as
\beq\label{critg}
\alpha(\mu) =
\frac{\al *}{1+W(\mu) }
\ ,
\eeq
where we introduced the shorthand notation $W(\mu)\equiv W[z(\mu)]$ with  $W[z]$ the Lambert function, and $z(\mu)$ given in \eq{z}. 
Together with \eq{critsurf}, 
the solution \eq{critg} fully describes the RG evolution of couplings along the separatrix to leading order in $\eps$.

\subsection{Characteristic energy scale and dimensional transmutation}\label{Scale}
The dynamics of the theory towards low energies is characterised by an energy scale $\Lambda_c$ where the RG flow displays a cross-over from UV fixed point scaling to either Gaussian scaling, or strong coupling, in the IR. The scale $\Lambda_c$ arises dynamically through the mechanism of ``dimensional transmutation'', and its r\^ole  is similar to that of $\Lambda_{\rm QCD}$ for QCD.
To identify the characteristic energy scale $\Lambda_c$,  we use\eq{dalpha2} or  \eq{solution2loop} to introduce the effective scaling exponent $\vartheta(\alpha)=\left.\partial_\alpha\beta_g\right|_{\rm sep}$ which reads
\beq\label{thetaB}
\vartheta(\alpha)=
-2B\,\alpha\,\left(1-\frac 32 \frac{\alpha}{\al *}\right)
\,.
\eeq
Close to the UV and IR fixed points, $\vartheta$ reduces to the conventional scaling exponents
\beq
\begin{array}{lcl}
\vartheta(\alpha\to 0)&=&-2\cdot B \cdot \alpha\to 0^+\\
\vartheta(\alpha\to \al *)&=&\ \ \ \ \ \ B\cdot \alpha_* <0
\end{array}
\eeq
showing that the gauge coupling is marginally attractive at the Gaussian IR fixed point $\vartheta\to 0^+$, and a relevant coupling at the interacting UV fixed point $\vartheta=B\cdot \alpha_*\propto - \eps^2$.  In between the fixed points, \eq{thetaB} changes sign at 
\beq\label{crossover}
\alpha=\frac 23\,{\alpha_*}\,,
\eeq
irrespective of the magnitude of $B$. The factor $\frac 23$ arises because \eq{thetaB} vanishes at the point where the one-loop and the two-loop contribution, proportional to $\alpha^2$ and $\alpha^3$, respectively, cancel out.  In fact, all four $\beta$-functions reach a maximum at \eq{crossover} along the separatrix. At this point, 
and with decreasing RG momentum scale,
 or decreasing $\alpha$, 
the RG flow leaves the basin of attraction of the UV fixed point and enters the domain of attraction of the IR fixed point. We define the characteristic energy scale $\Lambda_c$ 
as the scale where the UV-IR crossover happens. Using \eq{solution2loop} and \eq{crossover}, $\Lambda_c$ can be expressed in terms of the renormalisation group scale parameter $\mu$ and a small deviation of $\alpha$ from the UV fixed point $\delta \alpha=\al *-\alpha\ll \alpha_*$ as
\beq
\label{Lambdac}
\Lambda_c=c\cdot\mu\cdot\left|1-\frac{\alpha}{\al *}\right|^{\nu}\,,
\eeq
where we have suppressed terms of the order $|\delta \alpha|/ \alpha_*\ll 1$. The proportionality constant 
is entirely fixed by the universal physics in the UV and, here, given by $c_{\rm cross}=
\exp[(\ln 2-\s012)\nu]$ with $\nu=-1/\vartheta(\al *)$. Notice that the  scale $\Lambda_c$  is invariant under the renormalisation group,
\beq
\mu\,\frac{d}{d\mu}\,\Lambda_c=0\,.
\eeq
It arises  from a dimensionless quantity, the small deviation $|\delta\alpha(\mu)|/\al *\ll 1$ from the UV stable fixed point at asymptotically large scales $\mu\gg \Lambda_c$ through the phenomenon known as ``dimensional transmutation''. For $\delta\alpha<0$, and below the scale $\Lambda_c$, the theory is characterised by the infra-red free dynamics of weakly-interacting gluons, fermions and scalars, controlled by the Gaussian fixed point.  A similar characteristic energy is linked to  the UV finite trajectory emanating from the UV fixed point towards strong coupling when $\mu\to 0$ ($\delta\alpha>0$). The crossover away from the UV fixed point is characterised by the very same scale \eq{Lambdac}. Provided there is no strongly-coupled infrared fixed point we may estimate  the scale where strong coupling sets in from $\alpha_g(\mu=\Lambda_c)\approx 1$, finding \eq{Lambdac} with $c_{\rm strong}=\exp \nu$.

In the regime \eq{small} , the results \eq{Lambdac} for the characteristic energy scale and the crossover relation \eq{crossover} are exact non-perturbatively, and to all loop orders.\footnote{The result for the exact crossover relation \eq{crossover} and the corresponding crossover scale $\Lambda_c$ has already been used in \cite{Sannino:2014lxa} to explore asymptotically safe models of dark matter.}
 Once $\epsilon$ is no longer taken to be asymptotically small, higher loop terms or even non-perturbative corrections will modify the relation \eq{crossover}, and, consequently, the numerical coefficient $c$ in \eq{Lambdac}. 

For illustration of our results in this section, we display in Fig.~\ref{pRunAll} the renormalisation group running of all couplings along the UV-IR connecting separatrix. In Fig.~\ref{pRunAll}, we have expressed all couplings as functions of $1/z$, where  $z(\mu)$, defined in \eq{z} is related to the RG momentum scale.  Using the expression \eq{critg} for the running gauge coupling, and the expressions \eq{critsurf} for the running Yukawa and scalar couplings to leading order in $\epsilon$, we observe that all coupling display a characteristic crossover from the interacting UV fixed point to the free IR fixed point. The crossover takes place at the RG invariant scale \eq{Lambdac} where $z\approx 1$. We notice that $\al g$, $\al y$ and $\al h$ stay positive throughout, whereas  $\al v$ stays negative. We also observe that the sum of the quartic scalar coupings $\al h + \al v$ stays positive for all scales. This result plays a central role for the stability of the theory to which we turn next.

\section{Vacuum stability}
\label{CWA}
In the  analysis of the renormalisation group flow it was tacitly 
assumed that the vacuum of the scalar potential stays at the origin,
such that all global symmetries are preserved along the flow.
This hypothesis needs to be carefully scrutinised. In fact, we have already observed that the classical potential for one of the fixed points is unbounded from below and therefore the associated fixed point is not physical. We need to go beyond the classical stability for the other fixed point given that quantum corrections to the scalar potential can shift the minimum away from the origin. In perturbation theory this phenomenon is known as the Coleman-Weinberg mechanism~\cite{Coleman:1973jx}.
This is therefore a crucial consistency check for the quantum stability of the vacuum at the UV stable fixed point and the associated critical flow away from this fixed point.

\subsection{Classical moduli space}\label{moduli}
We start by recalling the stability analysis for potentials of the form
\begin{align}\label{V}
V = u\,\Tr\,(H ^\dagger H )^2  
+v\,(\Tr\,H ^\dagger H )^2 \,,
\end{align}
where $u$ and $v$ are real couplings, and the $N_F\times N_F$ matrices $H$ are invariant under $U(N_F)_L\times U(N_F)_R$ rotations. The main difference to the original analysis of Paterson \cite{Paterson:1980fc} is that the theory develops an interacting fixed point in the UV,
where $u$ and $v$ should take their fixed point values. At this level, stability of the vacuum requires that we can, at most, have a set of \emph{flat directions} along which we can find an infinite number of vacua degenerate with the one at the origin. The set of flat directions -- the classical moduli space -- is parametrized by 
the matrices $M_{ij}$ in field space with $\Tr M M^\dagger = 1$, 
$V(M) = 0$, and
$V(M + \delta M ) \geq 0$ for all $\delta M_{ij}$.
Since any classical field configuration $H_c$ can be diagonalized by a chiral \mbox{$U(N_F)_L\times U(N_F)_R$}  rotation, we can write $H_c = \text{diag}(h_1,\ldots,h_{N_F})$ without loss of generality. Consequently, \eq{V} becomes
\begin{align}\label{rotatedV0}
V= u \sum_{i=1}^{N_F} h_i^4 + v \left (\sum_{i=1}^{N_F} h_i^2 \right )^2 \,.
\end{align}
If $V$ vanishes for some $H_c$ it will vanish for any
multiple of $H_c$. It is thus sufficient to consider $V$ on the hypersphere
$\sum_i h_i^2 =1$. The stationary points of $V$ on this
hypersphere can be found by introducing a Lagrange multiplier 
$-2 \lambda (\sum_i h_i^2 -1)$ in \eq{rotatedV0} and solving the
stationary equation:
\begin{align}
\frac{\partial V}{\partial h_j} = 4 h_j \left [u h_j^2+v  - \lambda \right ] = 0 \quad{\rm (no\ sum)}\ .
\end{align}
The solutions are either
$h_j = 0$ or
$ h_j^2 = \frac{\lambda - v}{u}$,
showing that all nonzero $h_i$ must be equal at an extremum.
From the hypersphere constraint it follows that
for $n$ nonzero elements $h_i$, each of these must be related to $n$ as
$h_i^2 = 1/n$. Evidently $n$ may take values between $1$ and $N_F$.
Consequently, at an extremum the potential reads
$V_{\rm extr}= v + u/n$.
The value of $n$ is determined by requiring the extremum to
be a minimum. 
This can be realized in two different manners: 
Provided that $u>0$, then $V_{\rm extr}$ is minimal for $n=N_F$, and consequently
$M_{ij} = \delta_{ij}/\sqrt{N_F}$. 
Alternatively, provided that $u<0$, then $V_{\rm extr}$ is minimal for $n=1$, and consequently
$M_{ij} = \text{diag}(1,0,\ldots,0)$. 
We conclude that for $V$ to be bounded from below, 
the couplings are constraint to the parameter region
$\{u>0 \wedge v +u/N_F  \geq 0$ for $M_{ij}\propto \delta_{ij}\}$, or to the region $\{
u<0 \wedge v + u\geq 0$ for $M_{ij}\propto \delta_{i1}\}$. 
Flat directions are obtained if, in either case, the second inequality becomes saturated.

\subsection{Vacuum stability at UV fixed points}
Using the findings of Sec.~\ref{moduli}, we now turn to the UV fixed points detected in our model. In the conventions introduced in \eq{couplings}, and depending on the sign of the single-trace scalar coupling, the viable domains correspond to either of
\bea
\label{plus}
\al h>0&\quad{\rm and}\quad& \al h+\al v \geq0\,,\\
\label{minus}
{\rm or}\quad\al h<0&\quad{\rm and}\quad& \al h+\al v /N_F \geq0\,.
\eea
The stability analysis for all fixed points is summarised in Tab.~\ref{Tab1}. Both  FP${}_1$ and FP${}_2$ have $\al h>0$, but the condition \eq{plus} only holds true for FP${}_1$. At the partial fixed point FP${}_3$, we observe $\al h<0$, but $\al v>0$ displays a Landau pole towards high energies. However, in the strict infinite $N_F$ limit, $\al v$ drops out of \eq{minus} and consequently, \eq{minus} cannot be satisfied for any $\al h$. We conclude that of all fixed points within the perturbative regime, it is FP${}_1$ \eq{alphaNNLO} which corresponds to a stable scalar field potential, whose global minimum is located at the origin in field space (see Tab.~\ref{Tab1}). Furthermore, the potential has no flat directions provided that $\eps>0$. At finite $N_F$, however, additional stable solution of the type \eq{minus} may arise in the regime of large $\al v$. Presently this cannot be decided based on the approximations adopted here.

\begin{center}
\begin{table}
\begin{tabular}{c|c|c|c}
 \hline\hline
{{}\ \ fixed point\ \ }
&
{\ \ single trace coupling\ \ } &${}\quad$ stability condition${}\quad$ 
&${}\quad$ stability fullfilled${}\quad$ 
\\ \hline
FP${}_1$     
&$\ \  \alpha_{h}^*>0\  \ $
&$\ \  \alpha_{h}^*+\alpha_{v}^*\ge 0\  \ $
&
yes\\ 
 FP${}_2$     
&$\alpha_{h}^*>0$
&$\alpha_{h}^*+\alpha_{v}^*\ge 0$
& no
\\ 
 FP${}_3$     
&$\alpha_{h}^*<0$&$\alpha_{h}^*+\alpha_{v}(\mu)/N_F\ge  0$&
$
\begin{array}{rc}
{\rm infinite\ }N_F:\ &\ {\rm no}\\
{\rm finite\ }N_F:\ &\ {\rm possibly}
\end{array}
$
\\  \hline\hline
\end{tabular}
\caption{\label{Tab1}
 Classical moduli space. Full stability of the scalar potential at infinite $N_F$ is achieved at the UV fixed point \eq{alphaNNLO} of the gauge-Yukawa theory. Notice that a second domain of stability, related to the partial fixed point FP${}_3$, opens up for finite $N_F$  and sufficiently large $\al v$.}
\end{table} 
\end{center}

\subsection{Quantum moduli space}

We now investigate the quantum moduli space through the explicit computation of the quantum effective  potential {\it \`a la} Coleman-Weinberg \cite{Coleman:1973jx,Gildener:1976ih}. We are particularly interested in the UV safe trajectory which connects the UV fixed point with the Gaussian IR fixed point. 
 From the moduli-space analysis in the previous subsection we can assume, without loss of generality, that $H_{ij} \rightarrow \phi_c \delta_{ij}$ where the constant classical field $\phi_c$ takes the r\^ole of the expectation value of the renormalised quantum field. The quantum effective potential thereby becomes a function of the renormlasied field, the renormalisation group scale, and the running couplings. As such, it obeys an exact renormalisation group equation 
\ea{
\label{RG}
\left (
\mu_0\frac{\partial}{\partial \mu_0}
- {\gamma}(\al j)\, \phi_c\frac{\partial}{\partial \phi_c}
+ \sum_i {\beta_i}(\al j)  \frac{\partial}{\partial \al i}  \right ) V_{\rm eff}(\phi_c,\mu_0,\alpha_j ) 
= 0 \,.
}
Similar exact renormalisation group equations can be written down non-perturbatively for general effective potentials \cite{Litim:1994jd,Freire:2000sx}. Here, the functions $\beta_i$ are given in \eq{betag} -- \eq{betav}, and the function $\gamma=- \frac{1}{2}{d\ln Z}/{d\ln \mu}$ 
denotes the scalar field anomalous dimension given by
\beq
\label{gamma}
\gamma=\alpha_y
-\s032\left(\s0{11}{2}+\eps\right)\alpha_y^2
+\s052\alpha_y\alpha_g
+2\alpha_h^2 +{\cal O}(\alpha^3)
\eeq
up to the two-loop order in perturbation theory  \cite{Litim:2014uca}. To solve the exact renormalisation group equation \eq{RG}, we exploit that the potential is a pure quartic at the fixed point. By dimensional analysis the effective potential can then be written as 
\ea{
\label{Veffgeneral}
V_{\rm eff}(\phi_c;\mu_0,\al i) =   \lambda_{\rm eff}(\phi_c/\mu_0,\al i)\cdot\phi_c^4 
}
in the absence of further mass scales. The dimensionless prefactor  $\lambda_{\rm eff}(\phi_c/\mu_0,\al i)$  can only be a function of dimensionless parameters such as the dimensionless couplings, and the background field in units of the renormalization scale $\mu_0$. It obeys the exact renormalisation group equation
\ea{
\label{RGF}
\left (
\phi_c\frac{\partial}{\partial \phi_c} +4\, \bar{\gamma}(\al j) - \sum_i \bar{\beta_i}(\al j)   \frac{\partial}{\partial \al i} 
\right ) \lambda_{\rm eff}(\phi_c)
= 0 \ ,
}
whereby we have traded the $\mu\partial_\mu$-dependence for the $\phi\partial_{\phi}$-dependence. We have also introduced  the new functions
\beq
\label{bargamma}
\begin{array}{rcl}
\displaystyle
\bar{\beta_i}(\al i) &=& 
\displaystyle
\frac{\beta_i(\al i)}{1+\gamma(\al i)}\\[2.5ex]
\displaystyle
\bar{\gamma}(\al i) &=& 
\displaystyle
\frac{\gamma(\al i)}{1+\gamma(\al i)} \,,
\end{array}
\eeq
which allow us to translate information about the renormalization group $\beta$-functions of couplings into the field-dependence of the potential. Integrating the renormalization group equation for $\lambda_{\rm eff}(\phi_c)$ with a suitable boundary condition $\lambda_{\rm eff}(\phi_c=\mu_0)$ is then equivalent to a 
resummation of logarithms in the field.
Technically, this is done by defining new  running couplings $\bar{\al i}$ through the differential equations
\ea{
\label{barg}
\phi_c\frac{d \bar{\al i}}{d\phi_c} 
= \bar{\beta_i}(\bar{\al i}) \ , \quad \text{with} \quad 
\bar{\al i}(\phi_c=\mu_0) = \al i \ .
}
This allows us to write a general solution of \eq{RGF} in terms of the couplings $\bar{\al i}$ which solve \eq{barg}, as
\ea{
\label{lambdaeff}
\lambda_{\rm eff}(\phi_c)=\lambda(\phi_c)\,
\exp \left(-4\int_{\mu_0}^{\phi_c} \frac{d\mu}{\mu}\,\bar \gamma (\mu)\right)\,.
}
where  $\lambda(\phi_c)\equiv \lambda[\bar\alpha_i(\phi_c)]$ can be an \emph{arbitrary} functional of the functions $\bar{\al i}$ whose scale-evolution is determined via \eq{barg}. Its field-dependence solely arises implicitly through the dependence of  $\bar{\al i}$ on $\ln(\phi_c/\mu_0)$.
The functional form for $\lambda(\phi_c)$ can  be fixed through a matching with the fixed order perturbative expansion. 
To one-loop accuracy, and recalling the normalisation convention \eq{couplings}, the function $\lambda(\phi_c)$ reduces to the sum of the trace and double-trace scalar couplings,
\beq\label{lambdac}
\lambda(\phi_c)=4\pi^2[\bar{\al h}(\phi_c)+\bar{\al v}(\phi_c)]\,,
\eeq
where 
the substitution $\al i (\phi_c)\to \bar{\al i}(\phi_c)$ ensures that the leading logarithms are resummed. The validity of conventional perturbation theory requires the smallness of both $|\al i|\ll 1$ and $|\al i \ln (\phi_c/\mu_0)| \ll 1$. The resummation reduces the two constraints to  a single one, $| \bar{\al i}(\phi_c) | \ll 1$. Consequently, the  regime of small fields can now be probed even though the logarithms $\ln(\phi/\mu_0)$ becomes large in the limit of small fields.\footnote{An example  showing that the fixed order effective potential without resummation is unable to capture the physical properties of the potential near fixed points is given in \cite{Antipin:2012sm}.}

\begin{figure*}[t]
\begin{center}
\includegraphics[width=.7\textwidth]{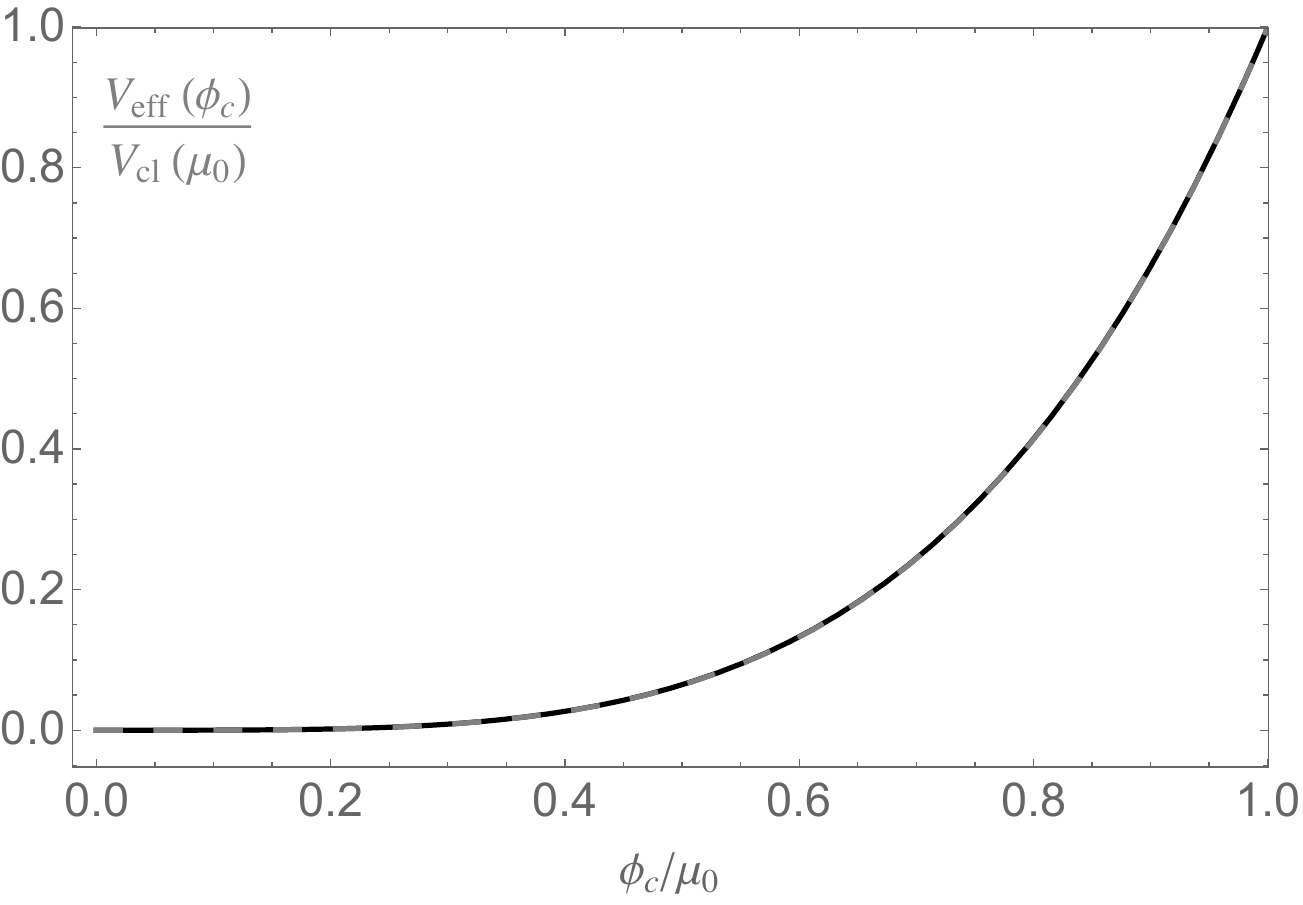}
\caption{\label{pCW} Stability of the quantum vacuum and absence of spontaneous symmetry breaking: shown is the Coleman-Weinberg effective potential $V_{\rm eff}(\phi_c)$ along the UV finite trajectory connecting the UV stable fixed point and the Gaussian IR fixed point, as a function of the constant classical field $\phi_c$ ($\eps=0.05$). The potential is normalised to the value of the classical fixed point potential at $\phi_c=\mu_0$.}
\label{pCW}
\end{center}
\end{figure*}

 \subsection{Analytical Coleman-Weinberg potential} \label{analytical}
We now derive  analytical expressions for the resummed quantum effective potential to NLO${}^\prime$ order in the approximation, the leading non-trivial order  in $\epsilon$ where the RG flow of scalar couplings becomes relevant. We can substitute $\bar\gamma$ by $\gamma$ and suppress the subleading $\gamma$-corrections in \eq{bargamma}. In this limit, the explicit solution \eq{critg} for the running gauge coupling  is valid, and the running of the scalar couplings along the separatrix strictly follows the running of the gauge coupling via \eq{separatrixLO}. 
To find an explicit expression for the exponential term in \eq{lambdaeff}, we recall from \eq{gamma}
that $\gamma=\alpha_y$ at one-loop.  By the virtue of \eq{separatrixLO}, this leads to $\gamma=D\,\alpha_g$ with $D=\frac{6}{13}$ on the trajectory of interest \eq{separatrixLO}.
The remaining integration is then performed analytically using the explicit form 
for the running gauge coupling \eq{critg}, together with \eq{W} and \eq{z}. The classical and the resummed quantum effective potentials then take the explicit form
\beq\label{Veff}
\begin{array}{rcl}
V_{\rm cl}(\phi_c)&=& 
\displaystyle
\lambda_*\, \phi^4_c\\[1ex]
V_{\rm eff}(\phi_c)&=& 
\displaystyle
\frac{V_{\rm cl}(\phi_c)}{1+W(\phi_c)}\left(\frac{W(\phi_c)}{W(\mu_0)}\right)^{-4D/B}
\end{array}
\eeq
where the positive quartic coupling is given by $\lambda_*=\eps\,\frac{16\pi^2}{19}(\sqrt{20+6\sqrt{23}}-\sqrt{23}-1)$ at the fixed point, see \eq{alphaNNLO}, and the exponent $-4D/B=18/(13\cdot \eps)>0$ is positive and parametrically large in the regime of interest \eq{small}. For the computation of $V_{\rm eff}$  we have adopted the UV boundary condition where $\lambda_{\rm eff}(\phi_c=\mu_0)$ is very close to $\lambda_*$ to ensure that couplings are in the close vicinity of the UV fixed point. This implies that 
$W(\mu_0)\ll 1$. The effective potential then reduces to the classical potential at the fixed point in the limit $\phi\to\mu_0$. Corrections due to the resummation of logarithms arise once $\phi_c<\mu_0$, and they can become sizeable once $\phi_c\ll\mu_0$. The tendency of these corrections can be understood from \eq{Veff}  as follows. With decreasing $\phi_c<\mu_0$, we have that $W(\phi_c)/W(\mu_0)\approx  (\mu_0/\phi_c)^{-B\al *}$. 
This leads to a tiny reduction $\propto (1-W(\mu_0) (\mu_0/\phi_c)^{-B\al *})$  of $V_{\rm eff}$ over $V_{\rm cl}$  due to the running of the gauge coupling, 
and to an enhancement 
$\propto  (\mu_0/\phi_c)^{(16\,\eps)/(19)}$ due to the anomalous dimension of the scalar field. Evidently, the enhancement  wins, meaning that  the resummed logarithmic corrections increase $V_{\rm eff}$ over $V_{\rm cl}$  with decreasing $\phi_c/\mu_0$.  We also note that the shape of the potential is entirely dominated by the vicinity to the UV fixed point. It is only for field values $\phi_c$ of the order of the scale $\Lambda_c$ \eq{Lambdac} and below where the Gaussian fixed point takes over the control of the resummation of logarithms.

For want of completeness, we now investigate the corrections to the effective potential \eq{Veff} due to the additional resummation of the scalar field anomalous dimension in \eq{bargamma}, \eq{barg}. Integrating \eq{barg} for $\bar \alpha_g$ using the two-loop result for $\beta_g$ and the one-loop expression for $\gamma$ leads to the explicit solution 
\beq\label{baralpha}
\bar\alpha(\phi_c)=\frac{\alpha_*}{1+\W(\phi_c)}
\eeq 
for the gauge coupling, with $\W(\phi_c)\equiv W[\bar z(\phi_c)]$, where $W$ is the the Lambert function and $\bar z(\mu)$ given by the expression 
\bea
\label{zbar}
\bar z&=&
\left(\frac{\mu_0}{\mu}\right)^{-B\cdot\alpha_*/(1+D\cdot\alpha_*)}
\left(\frac{\alpha_*}{\bar\alpha_0}-1\right)
\exp\left(\frac{\alpha_*}{\bar\alpha_0}-1\right)\,.
\eea
Notice that $z$ in \eq{z} is identical to $\bar z$ in \eq{zbar} after the replacement $B\alpha_*\to B\alpha_*/(1+D\,\alpha_*)$. The effect of the additional resummation, therefore, is that the running of the gauge coupling $\bar\alpha_g(\phi_c)$ from the UV to the IR is mildly accelerated over the running of $\alpha_g(\phi_c)$ at the same scale $\phi_c$. Since both couplings are monotonically decreasing functions along the UV-IR connecting trajectory, this implies that
$\bar\alpha_g(\phi_c)\le\alpha_g(\phi_c)$. Equality  holds only at the IR and UV fixed points, corresponding to the limits $\phi_c\to 0$ and $\phi_c\to \mu_0$ respectively. Furthermore, the relations \eq{critsurf} continue to hold true for the couplings $\bar\alpha_i$ because all functions $\bar\beta_i$ follow from $\beta_i$ through a rescaling by one and the same factor. For this reason, $\lambda(\phi_c)$ takes the same functional form given in \eq{lambdac}, with $\bar\alpha_h$ and $\bar\alpha_v$ expressed as functions of $\bar\alpha_g$ in \eq{baralpha}, \eq{zbar}. 
Similarly, the exponential factor in \eq{lambdaeff} can also be integrated analytically for $\bar\gamma$ given in \eq{bargamma}. The effective potential then takes the form
\beq\label{Veffbar}
V_{\rm eff}(\phi_c)=\frac{V_{\rm cl}(\phi_c)}{1+\W(\phi_c)}\left(\frac{\W(\phi_c)}{\W(\mu_0)}\right)^{-4\frac{D}{B}(1+D\cdot\alpha_*)^{-1} }
\left(\frac{1+D\,\alpha_*+\W(\phi_c)}{1+D\,\alpha_*+\W(\mu_0)}\right)^{-4\frac{D}{B}\frac{D\cdot\alpha_*}{1+D\cdot\alpha_*}}\,.
\eeq
Comparing \eq{Veffbar} to \eq{Veff}, and also \eq{z} with \eq{zbar}, we observe that the additional resummation has induced corrections of the order of $D\,\alpha_*$.  Since $D\,\alpha_*=\s04{19}\eps\ll 1$, these corrections are parametrically suppressed by $\eps$ in the regime \eq{small}, and, additionally, by a factor of $\s04{19}$. Quantitatively,  \eq{Veffbar} and \eq{Veff} are essentially indistinguishable for fields $\phi_c$ larger than the characteristic scale $\Lambda_c$ \eq{Lambdac} where  the underlying RG flow crosses over from UV to IR scaling. Once $\phi_c$ is of the order of $\Lambda_c$ and below, meaning $W, \W\approx \s012$ and above, the potential \eq{Veffbar} is enhanced over \eq{Veff} starting in the 5-10\% regime (for $\eps=0.05)$. We conclude that the  additional corrections due to $\gamma\to\bar \gamma$ and $\alpha_i\to \bar \alpha_i$  in \eq{bargamma}, \eq{barg} are minute and quantitatively negligible unless $\phi_c\ll \Lambda_c$, and we can stick to  the result \eq{Veff} for all technical purposes.

\begin{figure*}[t]
\begin{center}
\includegraphics[width=.7\textwidth]{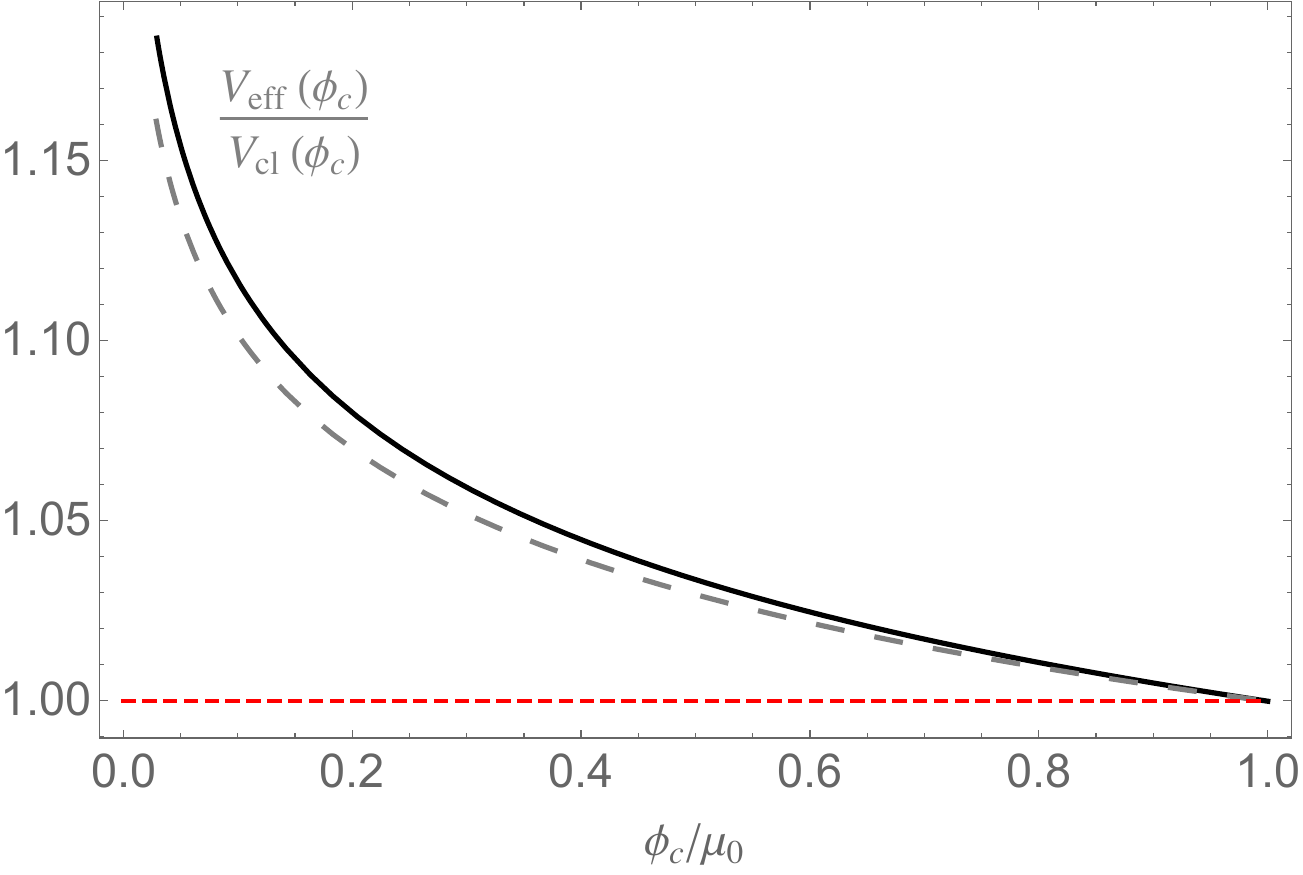}
\caption{\label{pCWnorm}  Enhancement of the quantum effective potential over the classical one: shown are the resummed quantum effective potentials $V_{\rm eff}(\phi_c)$ at different  orders in the approximation  (NLO${}^\prime$: dashed line, NNLO: full line) for $\epsilon=0.05$, and normalised to the classical potential $V_{\rm cl}(\phi_c)$ at the UV fixed point, see \eq{Veff}. At leading logarithmic accuracy, the enhancement $V_{\rm eff, NNLO}(\phi_c)\ge V_{{\rm eff, NLO}^\prime}(\phi_c)\ge V_{\rm cl}(\phi_c)$ is due to $\gamma_{{}_{\rm NNLO}}\ge\gamma_{{}_{\rm NLO}}\ge0$  (see main text).}
\label{pCompare}
\end{center}
\end{figure*}

 \subsection{Coleman-Weinberg potential and strong coupling}

As indicated earlier, two UV complete trajectories emanate from the UV stable fixed point. The first UV complete trajectory, towards weak coupling, connects the UV fixed point with the Gaussian IR fixed point. The cross-over to Gaussian behaviour takes place at the RG invariant scale \eq{Lambdac}. The second UV complete trajectory runs towards a regime of strong coupling in the IR. Again, the cross-over towards strong coupling takes places at a scale $\Lambda_c$ as in \eq{Lambdac}. Interestingly, the results of the previous subsection for the Coleman-Weinberg potential are applicable for either of these trajectories. The main difference between them is that the gauge coupling decreases in the former case, but grows in the latter. As we have discussed in Sec.\ref{analytical}, close to the UV fixed point  the running of the gauge coupling is too slow to leave an impact on the Coleman-Weinberg potential. Rather, it is the sign of the scalar field anomalous dimension which controls the logarithmic quantum corrections. To leading order, and for fields $\phi_c$ larger than the characteristic scale $\Lambda_c$, these corrections are largely independent  of whether the gauge coupling grows or decreases. Hence, with initial conditions $\al 0$ at $\mu_0$ on either side of, but very close to, $\al *$, the UV fixed point controls the logarithmic corrections and the corresponding effective potentials  become indistinguishable from each other. On the other hand, for much smaller fields, say of the order of  $\phi_c\approx \Lambda_c$, subleading differences between the potentials for weak and strong gauge coupling start to become visible. In this regime, perturbation theory is no longer a good approximation along the strong-coupling trajectory. For either of these trajectories, we conclude that 
 quantum effects neither destabilise the symmetric vacuum, nor do they lead to the spontaneous breaking of symmetry.

 \subsection{Numerical Coleman-Weinberg potential}
 We now turn to a full numerical evaluation of the quantum effective potential at NNLO accuracy,
 corresponding  to (3,2,1)-loop accuracy in the gauge, Yukawa, and scalar sector, respectively, using the $\beta$-functions as given in  \eq{betag} -- \eq{betav}.  Unlike the preceeding approximation, the RG flows at these orders obey Weyl consistency conditions. Concretely, for $\eps=0.05$, we determine the UV fixed point and the UV-IR connecting trajectory numerically for all couplings. These are then exploited to find the quantum effective potential \eq{Veffgeneral} by numerical integration of \eq{barg} and \eq{lambdaeff}. 

\begin{figure*}[t]
\begin{center}
\includegraphics[width=.77\textwidth]{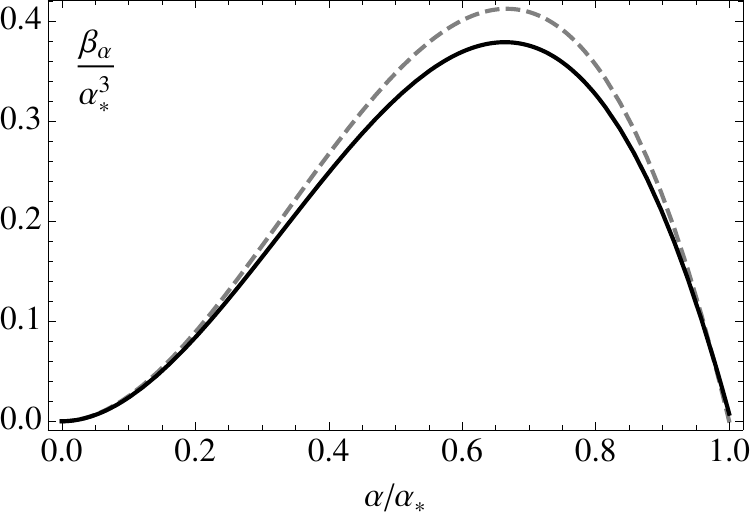}
\caption{\label{pSep}  Shown is the gauge $\beta$-function along the UV-safe trajectory at NLO (dashed line) and NNLO (full line) accuracy, with $\eps=0.05$. The main NNLO effect  consists in a slight reduction of the $\beta$-function once couplings are away from their fixed points.}
\label{pSepCompare}
\end{center}
\end{figure*}

Our findings are shown in Fig.~\ref{pCompare}  in comparison with the analytical results at NLO${}^\prime$ order \eq{Veff} and \eq{Veffbar}. As can be read off from  Fig.~\ref{pCompare}, with decreasing $\phi_c/\mu_0$, the effective potential $V_{\rm eff}$ at NNLO (full line) is enhanced over the NLO${}^\prime$ result (long dashed line). The red short dashed line indicates the result without resummation of logarithms. The slight enhancement of the NNLO result over the NLO${}^\prime$ result is understood as follows. The main  effect relates to the scalar anomalous dimension $\gamma=\al y$
whose value $\gamma_{\rm NNLO}$ is enhanced  over $\gamma_{\rm NLO}$ due to a slight shift in the fixed point value of $\al y$ from NLO${}^\prime$ to NNLO.  At $\eps=0.05$, the difference is approximately  20\%. In view of \eq{lambdaeff}, \eq{lambdac}, this leads to the relative change
\beq\label{ratio}
\frac{V_{\rm eff, NNLO}}{V_{{\rm eff, NLO}^\prime\ \ }}\approx
\frac{(\alpha_h^*+\alpha_v^*)_{\rm NNLO}} {(\alpha_ h^*+\alpha_v^*)_{{\rm NLO}^\prime\ \ }}
\left(\frac{\mu_0}{\phi_c}\right)^{4(\gamma_{\rm NNLO}-\gamma_{\rm NLO})}
\eeq
to leading logarithmic accuracy in $\phi_c/\mu_0$. The estimate \eq{ratio} is in very good quantitative agreement with the findings of  Fig.~\ref{pCompare}.
A secondary effect relates to  the running of the gauge coupling which at NNLO is slightly slower that at NLO${}^\prime$. This is evidenced in Fig.~\ref{pSepCompare}, and by the fact that the relevant scaling exponent has changed by roughly 10\% from $\vartheta_1\approx -0.0015$ at NLO${}^\prime$ to $\vartheta_1\approx -0.0014$ at NNLO. In principle this would lead to a small reduction of the ratio \eq{ratio}, which, however, is exponentially small and hence  of no relevance for the potential in the region considered here. We can therefore conclude that the enhancement \beq
V_{\rm cl}(\phi_c)\le  V_{{\rm eff, NLO}^\prime}(\phi_c)\le V_{\rm eff, NNLO}(\phi_c)
\eeq
of the effective potential for $\phi_c<\mu_0$ to leading logarithmic accuracy is largely due to the scalar anomalous dimension whose fixed point value
$0\le\gamma_{{}{\rm NLO}}\le \gamma_{{}{\rm NNLO}}$
changes from order to order in the approximation. Evidently, in this theory, quantum effects neither destabilise the symmetric vacuum, nor do they lead to the spontaneous breaking of symmetry.

Finally, we discuss our results in the light of perturbation theory and Weyl consistency conditions. The interacting UV fixed point arises for the first time at NLO' accuracy, see Tab.~\ref{Tab}. Stability of the vacuum  is observed,  persisting to finite values of  $\eps$. Stability  persists beyond NLO' accuracy, where non-universal parameters, such as the gauge three-loop coefficients or the Yukawa two-loop coefficients, come into play. This strengthens the view that the fixed point exists for finitely many fields.
Moreover, we observe that the stability of the vacuum is compatible with Weyl consistency conditions. The latter arise from formal considerations of  path integrals on curved backgrounds, and relate scheme-dependent orders in perturbation theory, e.g.~the NLO and the NNLO approximation. Weyl consistency conditions are {\it prima facie} insensitive to physical observables such as minima of  effective potentials. In this light it is noteworthy that Weyl consistency is  compatible  with vacuum stability.

  \section{Conclusions}
 \label{Conclusions}

Recently,  novel classes of fundamental four-dimensional quantum field theories with  non-Abelian gauge fields, fermions, and scalars have been discovered  whose high-energy behaviour is asymptotically safe, controlled by an exact interacting UV fixed point \cite{Litim:2014uca}.
Renormalisation group trajectories emanating from the fixed point relate to well-defined, finite, and predictive local quantum field theories at all energies, despite of the fact that asymptotic freedom is absent. The fixed point occurs parametrically close to the Gaussian and admits rigorous control within perturbation theory.

We have extended the study of \cite{Litim:2014uca} to establish that the vacuum of UV safe gauge-Yukawa theories 
is stable, classically and quantum-mechanically, even though asymptotic freedom is absent.
We also found that the main quantum corrections  to the effective potential  arise due to the  anomalous dimension of the scalars. Unlike in asymptotically free theories, here, the scalar anomalous dimension takes a non-vanishing value even at highest energies. 
The renormalisation group running of couplings away from the fixed point is a subleading effect for the effective potential, provided field values remain large compared to the characteristic energy scale $\Lambda_c$ \eq{Lambdac} of the theory. The absence of classically flat directions of the fixed point potential thus entails quantum stability.  Owing to the perturbative nature of the fixed point, we also determined the crossover relation \eq{crossover} exactly, including analytical expressions for all running couplings along the UV safe trajectories.

We have limited our investigation to massless theories in the Veneziano limit where 
the number of fields is  very large and UV interactions 
are weak. Continuity 
in the number of fields
 indicates that the vacuum remains stable even for finitely many fields as long as perturbation theory remains a good approximation. At strong coupling, 
 functional renormalisation can be used to access the Coleman-Weinberg potential non-perturbatively \cite{Litim:1994jd}.  Our proof of vacuum stability can straightforwardly be exported to other  gauge theories with interacting UV fixed points.  \\[4ex]

\centerline{\bf Acknowledgements}
This work is supported by the Science Technology and Facilities Council (STFC) [grant number ST/L000504/1]. The CP${}^3$-Origins centre is partially funded by the Danish National Research Foundation, grant number DNRF90.

\bibliographystyle{apsrev4-1}

\bibliography{UVsafeCW}

\end{document}